\documentclass[onecolumn]{aastex631}
\usepackage{mathtools}
\usepackage{amsmath}
\usepackage{soul}
\usepackage{lipsum}
\usepackage{gensymb}
\usepackage{rotating}

\newcommand\Rsun{\mathrm{R_\odot}}
\newcommand\Bo{\mathrm{B_0}}

\shorttitle{}
\shortauthors{Jaffarove et al.}

\begin{document}

\title{On the Relation between Source Region and In-situ Variability of Low $\delta B$ Solar Wind Streams}

\correspondingauthor{Kai Jaffarove}
\email{kai.j@berkeley.edu}

\author[0009-0008-6049-255X]{Kai Jaffarove}
\affiliation{Department of Physics, University of California, Berkeley, Berkeley, CA 94720-7300, USA}
\affiliation{Space Sciences Laboratory, University of California, Berkeley, CA 94720-7450, USA}

\author[0000-0002-8475-8606]{Tamar Ervin}
\affiliation{Department of Physics, University of California, Berkeley, Berkeley, CA 94720-7300, USA}
\affiliation{Space Sciences Laboratory, University of California, Berkeley, CA 94720-7450, USA}

\author[0000-0002-1989-3596]{Stuart D. Bale}
\affiliation{Department of Physics, University of California, Berkeley, Berkeley, CA 94720-7300, USA}
\affiliation{Space Sciences Laboratory, University of California, Berkeley, CA 94720-7450, USA}

\begin{abstract}
    Parker Solar Probe (PSP) observed a high speed stream near the Sun ($\sim 9.8 R_\odot$) in March 2025. As this stream was observed near the Sun, it allowed for an unparalleled opportunity to observe pristine, coronal hole wind with little evolution (expansion, stream interaction, etc) effects impacting its properties. Through an ensemble of magnetic connectivity analysis utilizing Potential Field Source Surface (PFSS) modeling and ballistic propagation, we make estimates of the footpoints, and corresponding source region parameters, associated with the stream. We then look at how the low variability regions (small gradient in the in-situ $B_r$) are related to the variability in the field strength at their source ($B_0$). We find that about half of these periods are associated with low source region variability. Lastly, we examine the low $\delta B / B$ periods between switchback patches and similarly find that they show statistically less variability in their associated $B_0$ value than their switchback patch counterparts. We believe that these results point to a solar source of these ``low-variability" periods, that warrants further investigation with composition diagnostics and more complex modeling techniques. 
\end{abstract}

\section{Introduction}

Spacecraft measurements have shown variability in solar wind in-situ parameters (magnetic field strength, velocity, density) for decades \citep[e.g.][]{Wilson-2021}. This variability has been shown to be a function of various factors such as the solar cycle \citep{Neugebauer-2002, McComas-2001, McComas-2008, Luhmann-2013} and solar source region \citep{Zhao-2017, Yardley-2024}, among other factors. The energization and magnetic topology at the solar source drives differences in charge state and elemental abundance ratios \citep{Landi-2012b} that are also often related to other bulk plasma parameters, like the wind speed, density, and temperature \citep[e.g.][]{Pagel-2004, Stansby-2020comp}. Near-Sun observations with the unique orbit of Parker Solar Probe \citep[PSP;][]{Fox-2016} have also shown the prevalence of microstream structures in the solar wind, which have been observed and studied since the 1990's \citep{Thieme-1990, Neugebauer-1995}. These are sub-structures within larger streams emerging from the same solar source.

The microstreams are found within fast wind streams and are typically characterized by fluctuations in the radial flow speed ($v_R$) above the background, and with rapid reversals of the magnetic field polarity ($\mathrm{sign(B_r)}$) known as switchbacks \citep{Raouafi-2023-Four}. These structures have been shown to be driven by interchange reconnection on supergranular scales \citep{Bale-2021, Bale-2023} and associated with quasi-periodic jets generated by reconnection in coronal holes \citep{Kumar-2023}. Other works have suggested in-situ generation mechanisms associated with the evolution of low-amplitude Alfv\'en waves present in the lower corona, leading to the development of circularly polarized, large amplitude waves through solar wind expansion \citep{Squire-2020, Mallet-2021}.

Between these microstream patches that contain the large amplitude waves (switchbacks), are quiescent regions \citep{DudokdeWit-2020, Short-2022, Short-2024}. These periods have been shown to have differences from their neighboring ``high-variability" periods that are filled with switchbacks. The quiescent periods show different spectral break points between the injection and inertial ranges in the turbulent cascade \citep{DudokdeWit-2020}, contain waves near the electron cyclotron frequency ($f_{ce}$) \citep{Short-2022}, and are more stable to kinetic instabilities than non-quiescent regions in the same region \citep{Short-2024}. \citet{Short-2024} proposes three hypotheses as to the origin of these regions: two related to their solar/coronal source and one related to an in-situ generation mechanism. They find strong evidence in favor of the solar/coronal source hypotheses as being more likely. 

During PSP Encounter 23, we observed one of the fastest streams near the Sun to date in the mission, which is full of microstream structures. These observations allow for the study of a pristine fast wind stream emerging from the center of a coronal hole. Various studies have used PSP (and other spacecraft) observations to study the source regions of the solar wind associated with what we observe in-situ \citep{Stansby-2018, Badman-2020, Badman-2023}. Recent works have shown that much of the wind we observe at PSP is slow (as it has not fully accelerated) and emerges from coronal holes (CHs) and their boundaries \citep{Stansby-2018, Bale-2019, DAmicis-2021}. These streams continue to accelerate as the wind expands, and it is likely that their evolution is highly impacted by interactions with neighboring stream structures \citep{Richardson-2018}. The close perihelion of PSP in its most recent Encounters with the Sun ($\sim 9.8 \Rsun$) allows us to study wind streams that are less impacted by evolution effects (expansion, stream interactions, transients) as well as streams below the Alfv\'en surface \citep{Badman-2025}. This specific stream provides a unique opportunity to study pristine fast wind (not regularly observed) very near the Sun ($9.8 - 11.5 \Rsun$).

In this work, we perform a connectivity analysis to make estimates of the footpoints of the fast wind stream. We then extract parameters of remote observations of the source and study how variability in the source region relates to variability observed in-situ. We specifically study ``low-variability" regions (regions with small gradients in their in-situ $B_r$ measurement), similar to the quiescent regions previously discussed. Lastly, we discuss further studies that could be carried out with more sophisticated modeling techniques and larger datasets.


\section{Modeling} \label{sec:modeling}
By combining models and our near-Sun PSP measurements, we can reconstruct estimates of the solar origins of in-situ measured wind streams to study how variability in the source drives what we see in-situ. Making the source to in-situ connection is vital for our global picture of the solar wind. Estimating this connectivity relies on our ability to model the coronal magnetic field, which is based on a series of complex steps (and assumptions). Input observations and parameter choices can also introduce uncertainty in these models, meaning that any resulting footpoints are truly an estimation of the connectivity, and many extrapolations need to be run to assess the stability of our models.

In this work, modeling is done via the classic two-step process of extrapolating the observed photospheric magnetic field using the Potential Field Source Surface \citep[PFSS;][]{Altschuler-1969} model and then propagating the spacecraft trajectory down to meet the model outer boundary (the source surface). In the PFSS model, we extrapolate an observation of the photospheric magnetic field out to a source surface ($R_{ss}$) where the modeled field is assumed to be purely radial. The PFSS model assume a magnetostatic corona ($\nabla \times \mathbf{B} = 0$), such that you solve $\nabla \cdot \mathbf{B} = \nabla \cdot ( - \nabla \psi) = \nabla^2 \psi = 0$ between the inner (photospheric field) and outer ($R_{ss}$) boundary conditions. This produces a modeled field out to $R_{ss}$, in our case 2.5$\Rsun$ \citep{Hoeksema-1984}.

The spacecraft trajectory at its varying heliocentric distance is then connected to the model's outer boundary by propagating the trajectory down to $R_{ss}$ \citep{Snyder-1966}. In the ballistic propagation, we assume a constant speed and do not account for solar wind acceleration. It has been shown that the error on the constant speed assumption is $\leq 10^\circ$ due to the fact that we do not account for coronal rotation, which is a counteracting force to the solar wind acceleration \citep{Nolte-1973}. There is a variety of ongoing work looking into the effects of different acceleration profiles on connectivity estimates \citep[e.g.][]{Dakeyo-2024}, which could be incorporated in future studies. 

In this work, we produce an ensemble of PFSS models (where $R_{ss} = 2.5 \Rsun$) that are initiated with different photospheric magnetograms (over the time period of interest). From these different extrapolations (36 of them), we can determine a mean and uncertainty on the footpoint estimates. We note that there are various other uncertainty sources in this two-step modeling pipeline \citep[e.g.][]{Ervin-2024c} that are not examined in this work, but could be included in the future. 

Once footpoints have been estimated, we extract source region properties for comparison with what we observe in-situ, similar to \citet{Ervin-2024a, Ervin-2024b, Ervin-2024c}. Using remote observations of the photospheric field, we extract the magnetic field strength at the estimated footpoint ($\Bo$), which is an estimate of the footpoint field strength where the wind originated. We can also compute an expansion factor from our PFSS extrapolation which describes how expanded the modeled field line is relative to spherical ($1/r^2$) expansion. 

\begin{figure}
    \centering
    \includegraphics[width=\linewidth]{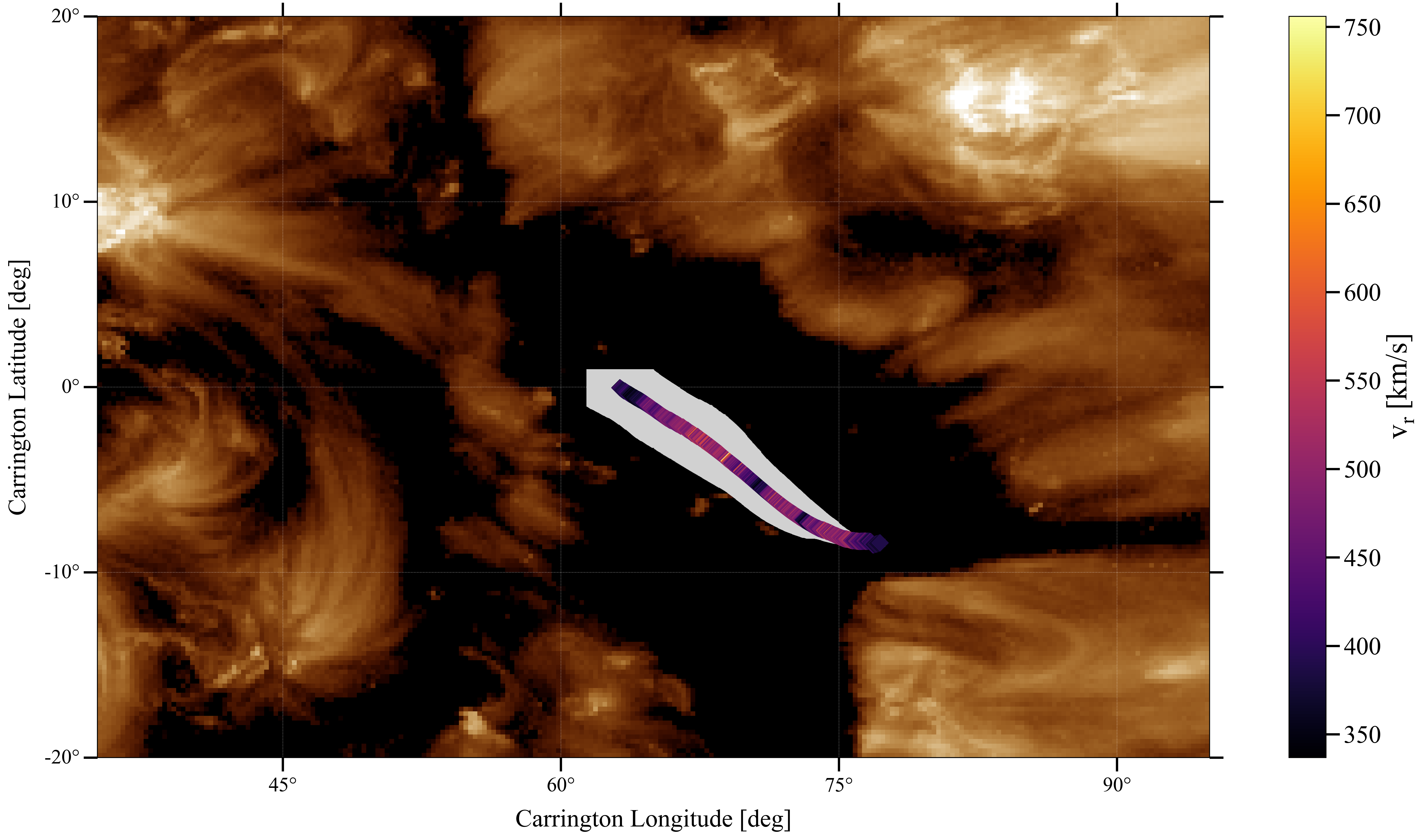}
    \caption{Estimated footpoints (PFSS + backmapping) colored by the measured in-situ proton radial velocity ($v_R$) plotted on a 193{\AA} SDO/AIA observation. The light gray shading shows the $1 \sigma$ uncertainty in the modeled footpoints associated with the ensemble of potential field extrapolations from different input magnetograms.}
    \label{fig:aia-mapping}
\end{figure}

In Figure~\ref{fig:aia-mapping}, we show the mean and uncertainty (from the different extrapolations) of the estimated footpoints for the stream of interest overplotted on an AIA EUV (193 {\AA}) observation of the source region. We see that even through varying the input magnetogram, we have stability in our source region, indicating trust in our modeling results. 

\section{Observations} \label{sec:observations}

PSP is an inner heliospheric spacecraft making high measurements of fields and particles to study the dynamics and structure of the source regions of the solar wind \citep{Fox-2016}. In this study, we use proton (partial) velocity moments (at 3.5 sec cadence) from the ion electrostatic analyzer \citep[SPANi;][]{Livi-2022} on board the SWEAP suite \citep{Kasper-2016}. While the SPANi instrument has an obstructed field of view (FOV) due to the instrument and spacecraft design, this FOV effect has a negligible impact on the calculated velocity moment \citep{Livi-2022}. From the FIELDS suite, we use the 4 samples / cycle fluxgate magnetometer (FGM) 3 axis magnetic field measurements \citep{Bale-2016}. We use the radial-tangential-normal (RTN) coordinate system for both the velocity and magnetic field measurements in this work, and downsample the FGM measurements to the 3.5 sec SPANi cadence when merged measurements are needed. 

\begin{figure}[ht!]
    \centering
    \includegraphics[width=\linewidth]{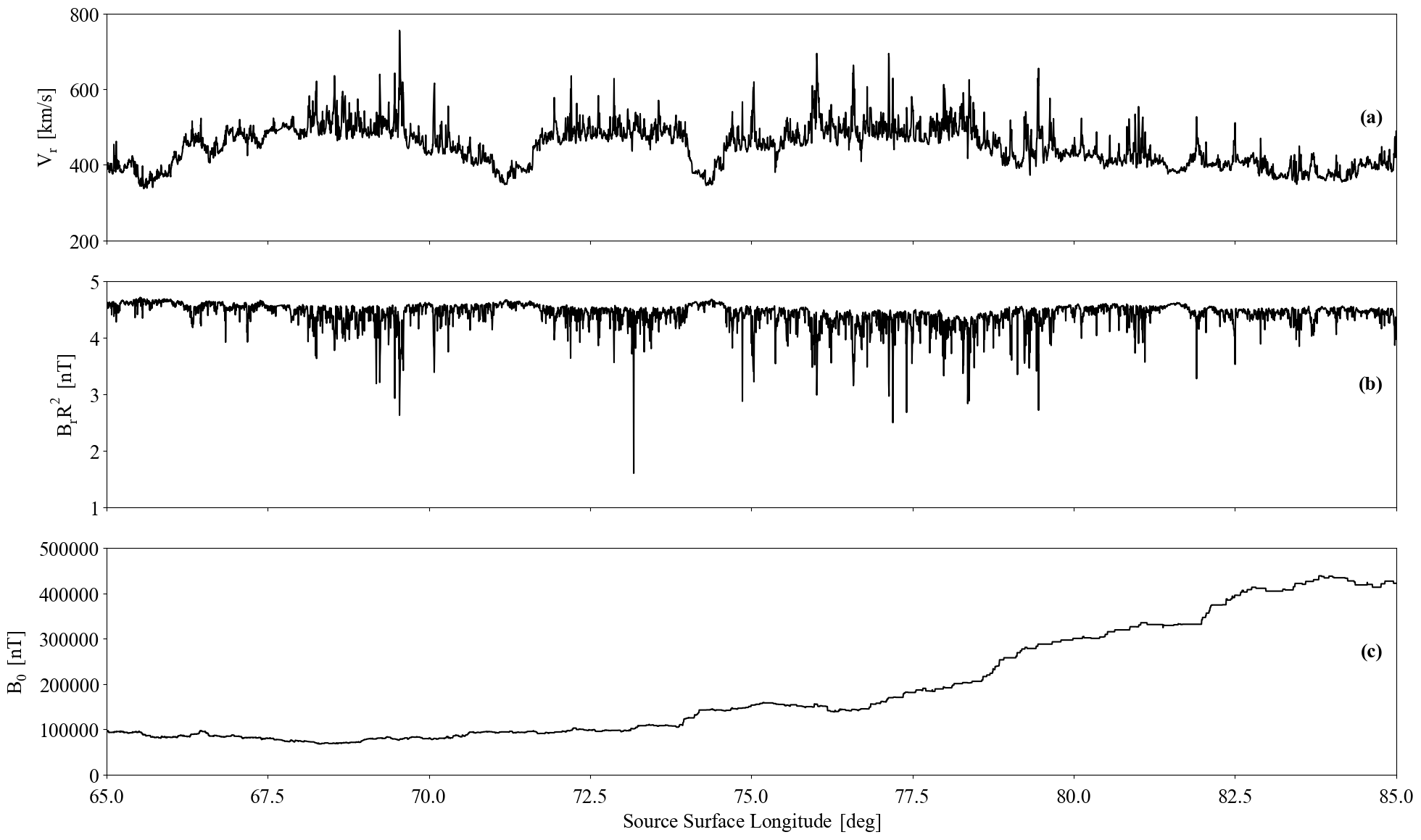}
    \caption{Overview of observations and modeling results for the Encounter 23 fast stream (2025-03-23 00:29:02 to 04:38:35). (a) the proton radial velocity (partial moment) from SWEAP/SPANi (b) the scaled radial magnetic field component (FIELDS/FGM) and (c) the footpoint field strength derived from combining modeling results and remote observations of the photospheric field (see Section~\ref{sec:modeling}).}
    \label{fig:observations}
\end{figure}

In Figure~\ref{fig:observations} we show the velocity and magnetic field measurements of interest. We examine a subsection of the fast solar wind (FSW) stream observed by PSP during Encounter 23 that spanned from 2025-03-23 00:29:02 to 04:38:35 ($9.8 - 11.5 \Rsun$). In panel (a), we show the SPANi proton radial velocity ($v_r$), and panel (b) shows the scaled radial magnetic field ($B_r R^2$). Panel (c) shows the footpoint field strength $\Bo$ (see Section~\ref{sec:modeling}), a measure of the estimated field strength at the source region. These values are plotted against the source surface longitude, calculated by projecting the spacecraft trajectory (assuming a constant speed) down to 2.5 $\Rsun$ assuming a Parker spiral (see Section~\ref{sec:modeling}). This converts a time series (as measured by the spacecraft) of the associated spacecraft position into a spatial series projected onto the source surface. 

Similar to previous works \citep[e.g.][]{Bale-2023}, we see distinct microstream structure within the fast wind stream. These ``microstreams” can be seen in both the (a) velocity measurements and (b) magnetic field  with a patch-like structure of switchbacks. Between the microstream, we observe periods of low variability in the magnetic and velocity field (e.g. quiescent regions \citep{Short-2024}). In panel (c), we show the footpoint magnetic field strength as derived in Section~\ref{sec:modeling}. 



\section{Discussion}  \label{sec:discussion}
\begin{figure}
    \centering
    \includegraphics[width=\linewidth]{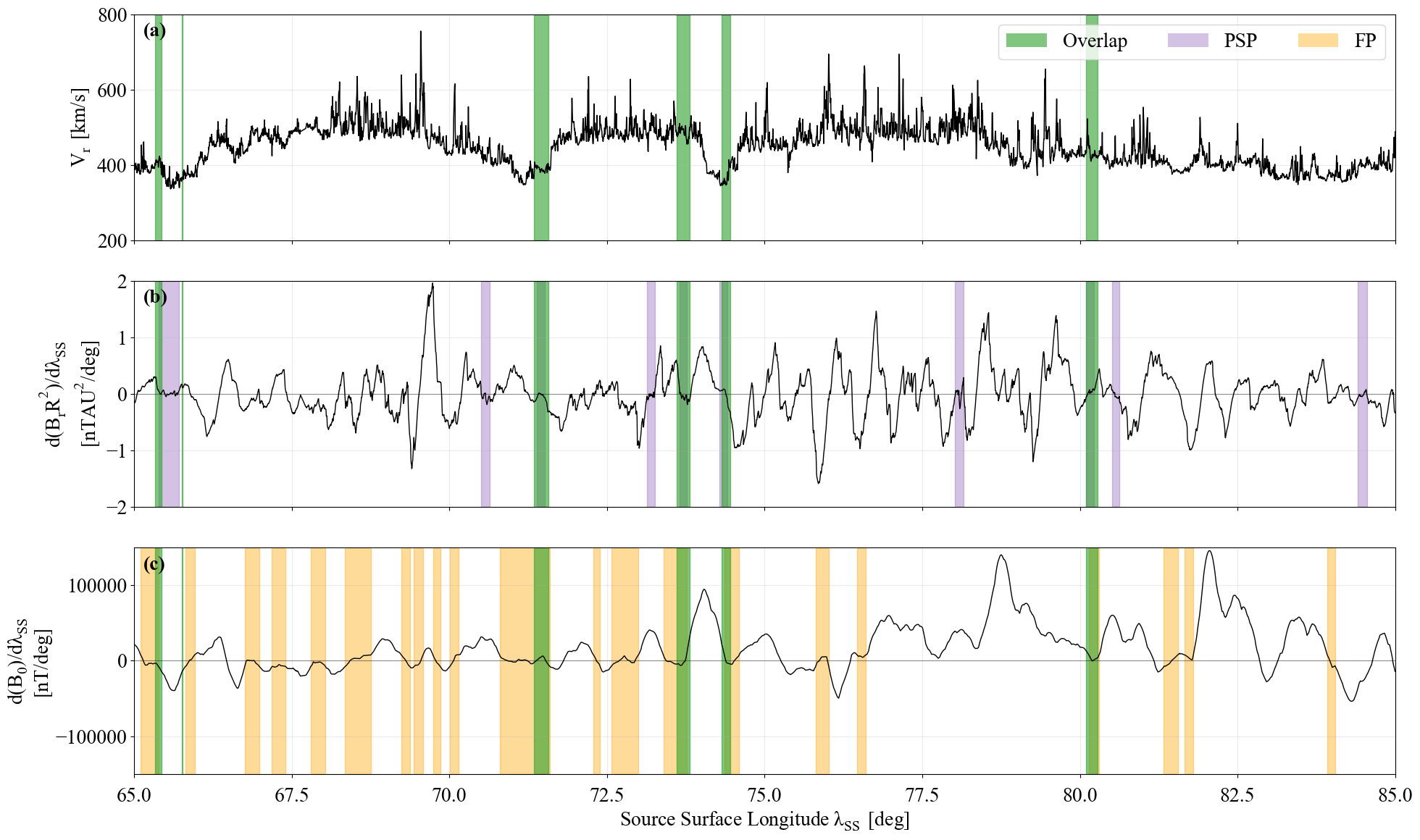}
    \caption{We calculate gradients of the magnetic field with respect to the source surface longitude at both (b) PSP and (c) the footpoints. Given certain selection parameters to determine low-variability regions, we identify all streams that satisfy these at PSP and the footpoints as seen in (b) purple and (c) orange respectively. If the longitudinal spans of these selected regions agree between PSP and the footpoints, we highlight them in green in all three panels. The top panel (a) shows the radial velocity at PSP, with the coinciding green regions projected onto it.}
    \label{fig:gradients-quiet}
\end{figure}

Now that we have estimates of our footpoint locations and have extracted footpoint field strength values from this, we move on to compare how the $\Bo$ variability relates to in-situ variability. In Figure~\ref{fig:gradients-quiet} we show a calculation of the spatial gradient of the in-situ magnetic field (Figure~\ref{fig:observations}(b)) and the source region $\Bo$ (Figure~\ref{fig:observations}(c)). These gradients are calculated as $dB_i / d \lambda_{ss}$ where $i$ is either $r$ for the in-situ measurement or $0$ for the photospheric field. Gradients are calculated using a half-width degree window of 0.2 degrees, corresponding to a 10-minute averaging window. 

Stream selection is determined by the magnetic field gradients at PSP and the photosphere as seen in Figure~\ref{fig:gradients-quiet}, checking whether they agree on certain ranges of longitudes. For regions of low magnetic variability i.e. the bottom 20\% of data, we highlight all of the identified streams at PSP in purple and the streams at the photosphere in orange. To determine the strictness of stream selection, we include a 0.05 source surface longitude buffer corresponding to the ballistic propagation error of fast streams \citep{Ervin-2024c}. We use a 0.2 half-width degree window corresponding to a 10-minute rolling window gradient calculation, and a 0.1 degree minimum width in order to consider a set of points as a stream, corresponding to 5 minutes. If streams in both regions agree for a given longitudinal range, we highlight that range in green in all three panels in Figure~\ref{fig:gradients-quiet}. 

From Figure~\ref{fig:gradients-quiet}, we can again qualitatively see that large gradients in (b) $B_r$ (large fluctuations) are associated with large gradients in (c) $\Bo$, while the low-variability regions (small $B_r$ gradient) are also associated with relatively stable values in $\Bo$. It is important to note that the projection onto the source surface makes various assumptions (e.g. constant wind speed, Parker spiral propagation) and thus we don’t expect an exact relationship between changes in the two quantities. 


To quantify how often the spacecraft measurements of low-variability magnetic field measurements agree (pair up) with the photospheric magnetic footpoint strength, we can simply calculate the proportion of paired streams relative to the total number of low-variability streams selected at PSP. In Figure~\ref{fig:gradients-quiet}, we find that 209 of the 396 samples, or 52.8\%, are considered to be low-variability streams at PSP, indicating that roughly half of the longitudinal regions considered at PSP coincide with low-variability footpoint conditions. This proportion suggests that a substantial fraction of the in-situ low-variability regions are often associated with low-variability at the photosphere, potentially indicating a connecting between source region dynamics and in-situ variability.

\begin{figure}
    \centering
    \includegraphics[width=\linewidth]{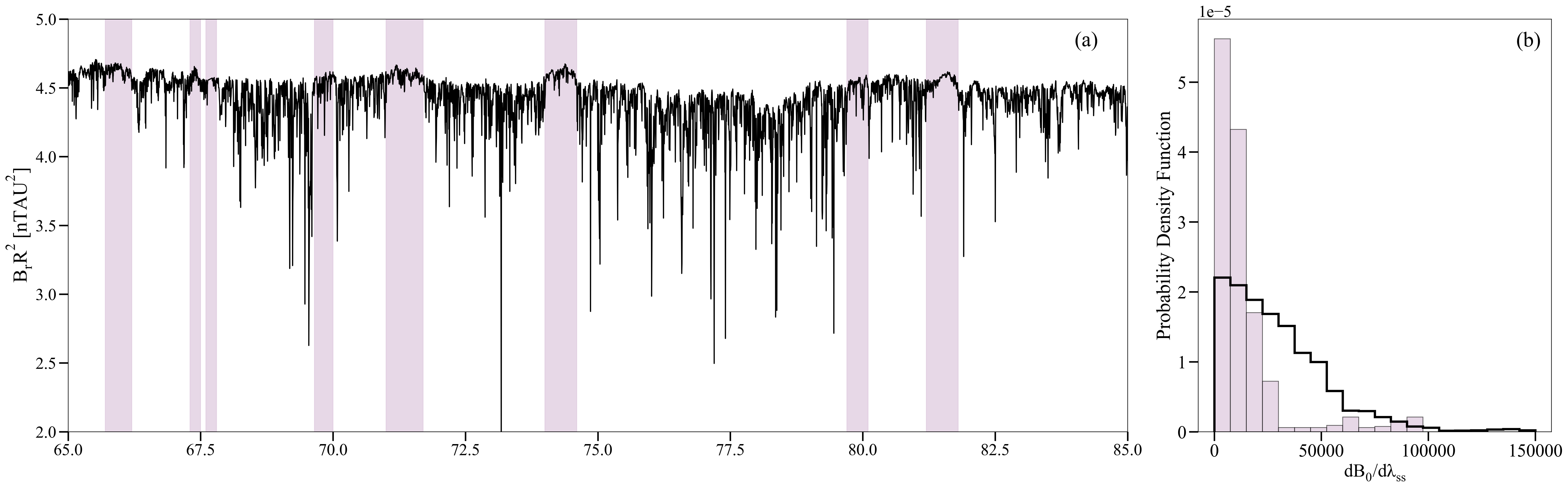}
    \caption{We manually identify regions where $\delta B / \langle |B| \rangle$ is small ($\leq 0.2$) to study the gradient in the source region magnetic field footpoint strength. (a) The scaled radial magnetic field highlighting our identified regions. (b) Probability density functions comparing changes in $\Bo$ between our identified regions (purple) and the rest of the observations (black).}
    \label{fig:source-variability}
\end{figure}

In Figure~\ref{fig:source-variability}, we manually identify low-variability regions to study how their footpoint gradient ($dB_0 / d \lambda_{ss}$) compares with that of the other periods. We find that $d\Bo$ is statistically smaller for the low-variability regions than for the full interval. This potentially supports ideas put forth by \citet{Short-2024} low-variability periods are associated with dynamics at the solar source.

We also look at the expansion factor of the low-variability regions compared to the full interval. We do not find any statistical difference in the magnitude of the expansion factor between the low-variability periods and the full interval. This should be further investigated with higher resolution magnetograms as input to the PFSS model and with conjunction studies looking at microstream expansion through the inner heliosphere. 

\section{Conclusion}  \label{sec:conclusion}
Through our solar wind connectivity analysis of this high-speed stream observed by Parker Solar Probe in Encounter 23, we are able to make estimates of source region parameters associated with in-situ observations. This allows us to directly study the relationship between the two and how changes at the source impact what the spacecraft observes. We find that source region variability such as changes in the expansion factor and footpoint field strength, is tied to the structure we observe in-situ. We note that there are various limitations to our analysis, specifically that our models only provide footpoint estimates (and they have some uncertainty), and thus we do not expect one-to-one correlation between source changes and in-situ changes. However, the relationship we do observe continues to grow our understanding and the idea that source changes are directly tied to in-situ variability.

Understanding how solar source variability drives in-situ variability is not only important for a complete picture of the heliosphere, but also for building robust space weather modeling capabilities. To understand what we expect to see at Earth (1 AU) distances, we must understand how the variability at the source drives the near-Sun wind. It is important to note that source variability is not the only driver of in-situ variability. Stream-stream interactions, interactions with transient structures, shear between fast and slow streams, and the solar cycle all lead to the solar wind variability we observe in-situ. However, with near-Sun observations from PSP, we can more closely examine the impact of source region variability. Future work looking at the evolution of this stream through the inner heliosphere could study the impact of expansion, plasma interactions, and turbulence on fast wind evolution.

Improvements in our modeling capabilities, alongside higher resolution remote observations, will allow us to have more precision in these sorts of studies. Potential field models do not capture any time-varying topological changes (they are steady state), nor can they account for eruptive events that are important for shaping the inner heliosphere and solar wind (especially at solar maximum). Using time-dependent models with more sophisticated physics \citep[e.g.][]{Downs-2025} could help address this issue. Global coverage of the solar magnetic field to full 3d extrapolations can be carried out with magnetograms that do not depend on flux transport algorithms to reconstruct the far side of the Sun are also needed to model streams observed by far-side spacecraft.

Elemental and charge state abundance ratios can trace the energization and the associated topological structure at the solar source. These heavy ion ratios are often frozen-in in the low corona providing in-situ metrics of coronal processes, however, measurements of heavy ions in the ambient solar wind are lacking. New missions that include time-of-flight composition measurements in the near-Sun environment could help disentangle the relation between source and in-situ variability.

\bibliography{ms}{}
\bibliographystyle{aasjournal}

\section{Acknowledgements} \label{sec: acknowledgements}
The FIELDS and SWEAP experiments on the Parker spacecraft was designed and developed under NASA contract NNN06AA01C. We acknowledge the NASA Parker Solar Probe Mission and the SWEAP team led by M. Stevens for use of data. TE acknowledges funding from The Chuck Lorre Family Foundation Big Bang Theory Graduate Fellowship. 

\end{document}